\documentclass[aps,pra,twocolumn,amsmath,amssymb,showpacs,floatfix,longbibliography]{revtex4-1}

\usepackage{amsmath}
\usepackage{txfonts}
\usepackage{microtype}
\usepackage{graphicx}
\usepackage{color}
\usepackage{ulem}
\usepackage{bm}

\begin{document}
\title{ Sub-barrier pathways to Freeman resonances}

\author{Michael Klaiber}\email{klaiber@mpi-hd.mpg.de}
\affiliation{Max-Planck-Institut f\"{u}r Kernphysik, Saupfercheckweg 1,
	69117 Heidelberg, Germany}
\author{Karen Z. Hatsagortsyan}\email{k.hatsagortsyan@mpi-hd.mpg.de}
\affiliation{Max-Planck-Institut f\"{u}r Kernphysik, Saupfercheckweg 1,
	69117 Heidelberg, Germany}
\author{Christoph H. Keitel}
\affiliation{Max-Planck-Institut f\"{u}r Kernphysik, Saupfercheckweg 1,
	69117 Heidelberg, Germany}

\date{\today}

\begin{abstract}

The problem of Freeman resonances [R. R. Freeman \textit{et al.}, Phys. Rev. Lett. \textbf{59}, 1092 (1987)] when strong field ionization is enhanced due to transient population of  excited states during the ionization, is revisited. An intuitive model is put forward which explains the mechanism of intermediate population of excited states during  nonadiabatic tunneling ionization via the under-the-barrier recollision and recombination. The theoretical model is based on  perturbative strong-field approximation (SFA), where the sub-barrier bound-continuum-bound pathway is described in the second order SFA, while the further ionization from the excited state by an additional perturbative step. The enhancement of ionization is shown to arise due to constructive interference of contributions into the excitation amplitudes originating from different laser cycles. The applied model provides an intuitive understanding of
the electron dynamics during a Freeman resonance in strong field ionization, as well as means of enhancing the process and possible applications to related processes.

\end{abstract}

\date{\today}

\maketitle

\section{Introduction}

 The enhancement of strong field ionization due to transient excitation of  Stark-shifted bound states is well-known from experiments in multiphoton regime of ionization and is termed as Freeman resonances \cite{Freeman_1987,Agostini_1989,Freeman_1991,Mevel_1993,Nandor_1999,Wiehle_2003,Morishita_2007,Wang_2009f,
 Potvliege_2009,Lu_2015,Hart_2016,Stammer_2020,Chetty_2020}. It is assumed that the excitation at Freeman resonances happens due to a bound-bound multiphoton transition,  when the electron wave function during transition is localized within the binding potential. With increased laser intensity, the tunneling through the laser suppressed Coulomb barrier becomes dominant and the bound electron moves from the ground state immediately to the continuum. Strong field approximation (SFA) \citep{Keldysh_1965,Faisal_1973,Reiss_1980} describes successfully direct strong field ionization in tunneling and multiphoton regimes as well as in the intermediate nonadiabatic regime \cite{Yudin_2001b,Ivanov_2005}, when the electron gains energy during the tunneling \cite{Klaiber_2015}. The quantum orbit picture \cite{Salieres_2001,Becker_2002} which stems from the SFA description, applying saddle-point approximation (SPA) in calculation of integrals in S-matrix amplitude, provides intuitive understanding of strong field ionization processes. Can the quantum orbit picture  be extended to interpret the electron dynamics at Freeman resonances?

In the tunneling regime the atom excitation due to bound-bound transitions is not probable, because it is overwhelmed by electron tunneling into the continuum. In the nonadiabatic tunneling  excitations can happen only when the electron revisits the atomic core, i.e. at  recollisions. However, the common recollisions via excursion in the real continuum \cite{Corkum_1993}, are accompanied by a large spreading of the electron wave packet, which reduces significantly the recollision probability. Recently, it has been recognized  that recollision can happen also within the sub-barrier dynamics during tunneling \cite{Klaiber_2018}. The latter may contribute to the electron transition to the excited state as long as the electron gains sufficient energy during the nonadiabatic tunneling.

\begin{figure}[b]
   \begin{center}
\includegraphics[width=0.5\textwidth]{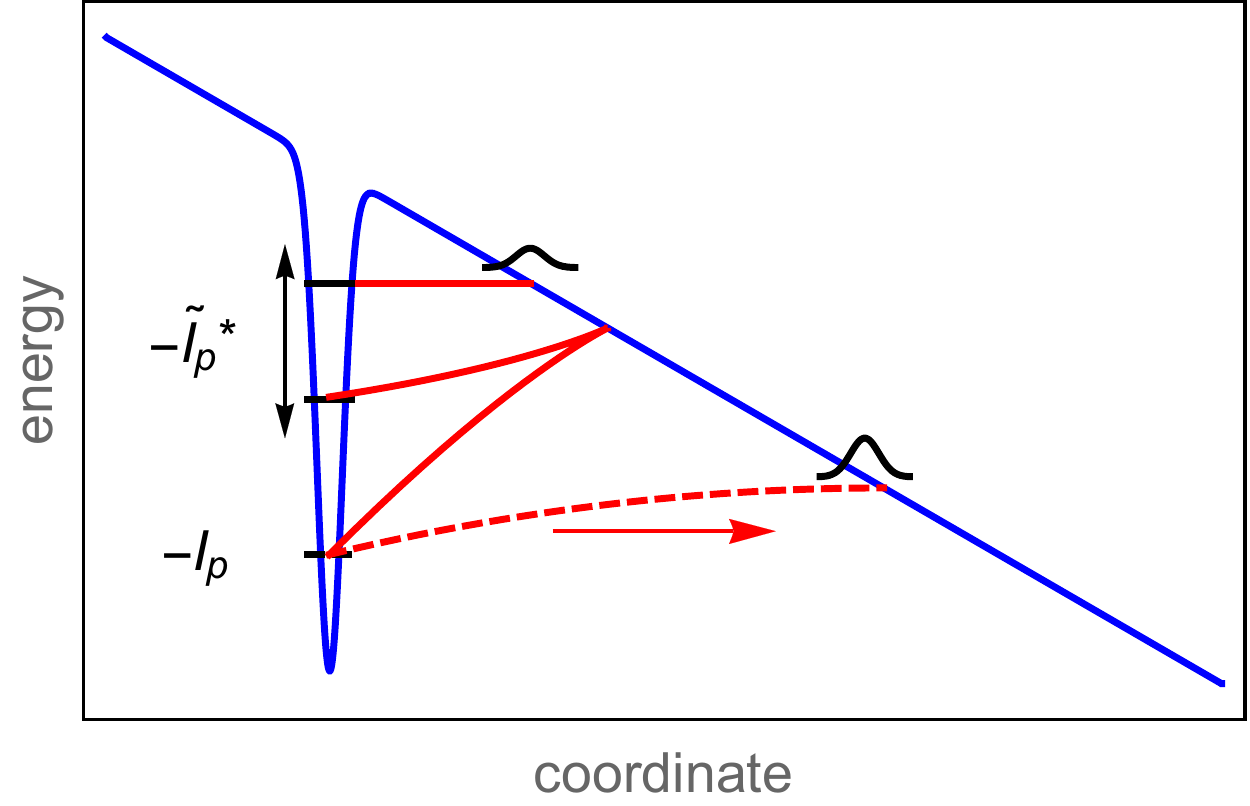}
  \caption{The scheme of the resonantly enhanced  nonadiabatic tunneling (Freeman resonance): the sub-barrier recollision (described by a quantum trajectory propagating from the bound state up to the barrier surface, reflected, and tunneled back to the core) may yield to recombination to the Stark-shifted excited state of the atom, which is followed by further ionization. The electron energy  in the excited state  is up-lifted due to the laser dressing during dwelling in the excited state before further ionization. Dashed line shows the path of the direct nonadiabatic ionization. }
         \label{scheme}
    \end{center}
  \end{figure}

In this paper we develop a theory for Freeman resonances in the nonadiabatic tunneling regime which is based on the concept of the under-the-barrier recollision. We employ SFA, treating the recollision with the atomic core within a perturbative approach. The resonant channel of ionization is described within the next-order perturbation of SFA. The given description allows for an interpretation of the process as taking place via sub-barrier recollision with increasing energy in the nonadiabatic regime and transition to the excited state, with further ionization after some time delay, see the interaction scheme in Fig.~\ref{scheme}. The proposed model provides a physical  explanation via quantum orbit picture for the resonantly enhanced  strong field ionization involving  excited states at a Freeman resonance.

The paper is organized as follows. The theoretical model is described in Sec.~\ref{sec:TM}. The half-cycle and multi-cyle contributions to the Freeman resonances are discussed in Secs.~\ref{sec:half-cycle} and \ref{sec:multi-cycle}, respectively. The photoelectron spectra within the present model are presented in Sec.~\ref{sec:spectra}, and our conclusion is given in Sec.~\ref{sec:conclusion}.

\section{Theoretical model}\label{sec:TM}

Our main aim is to provide an intuitive picture within the new scheme for Freeman resonances. For this purpose it is important to have analytical theory and, therefore, we illustrate the new scheme  in a simple and transparent one dimensional (1D) model.  We expect the picture to hold also in 3D, because the under-the-barrier recollision is virtually one dimensional along the parabolic coordinate even in the full 3D consideration.
The ionization dynamics of an atom in a strong laser field is described by the  Hamiltonian
\begin{eqnarray}
  H=\hat{p}^2/2+V(x)+ H_I(t),
  \label{Hamiltonian}
\end{eqnarray}
where $\hat{p}$ is the momentum operator, $V(x)$ is the potential of the atomic core, and $H_I(t)=xE(t)$ is the laser-electron interaction Hamiltonian, with the laser electric field $E(t)$.

The theoretical treatment is based on SFA. We begin with the exact ionization amplitude $m(p)$ for the photoelectron with a final momentum $p$:
 \begin{eqnarray}
m(p)=-i \int dt \langle \psi^V_{p}(t_f)|U(t_f,t)H_I(t)|\phi(t)\rangle\label{exact},
\end{eqnarray}
 where $U(t_f,t)$ is the exact time-evolution operator (TEO), with an asymptotic time $t_f$, $\psi^V_p(x,t)=\frac{1}{\sqrt{2\pi}}\exp[i(p+A(t))x+i S(t)]$ is the Volkov wave function \cite{Volkov_1935}, with the contracted classical  action $S(t)=\int_t^\infty ds[p+A(s)]^2/2$,  and $\phi(x,t)=\phi(x)\exp(i I_pt)$ is the wave function of the  atomic bound state, with the ionization potential $I_p$, $\kappa=\sqrt{2I_p}$ is the atomic momentum. The linearly polarized laser pulse is described by the vector potential $A(t)= (E_0/\omega)f(t) \sin(\omega t)$, with the field amplitude $E_0 $, the frequency $\omega$, $E(t)=-\frac{\partial A}{\partial t}$, and the slowly varying pulse envelope $f(t)$.

We describe the strong field ionization via resonant excitation during nonadiabatic tunneling. This pathway includes an under-the-barrier recollision due to which a transition to the excited state happens, from where the electron is readily ionized via tunneling or an over-the-barrier passage. To model the described pathway, in Eq.~(\ref{exact}) we need to  approximate  the exact TEO $U(t',t'')$, which is designed to describe  the laser driven sub-barrier dynamics of the electron, including an intermediate revisiting the atomic core.
For this reason we represent the exact TEO  symbolically as follows
\begin{eqnarray}
U(t',t'')=\sum_n |\tilde{\phi}_n(t')\rangle\langle \tilde{\phi}_n(t'')|,\label{U2}
\end{eqnarray}
with the sum running over the exact basis set $|\tilde{\phi}_n(t)\rangle$, representing the exact solutions of the Schr\"odinger equation in the laser  and the  atomic potential fields.
As $U(t',t'')$ recounts the dynamics via direct ionization and through the  laser-dressed excited state, we extend the sum in Eq.~(\ref{U2}) over continuum states and  bound states. Taking into account that  for the direct ionization the influence of the potential is negligible and during the resonance only one excited state $|\tilde{\phi}^*(t)\rangle$ is  important in the sum of Eq.~(\ref{U2}), the one which has an energy that fits to the energy of the recolliding electron, we approximate:
\begin{eqnarray}
U(t',t'')\approx |\tilde{\phi}^*(t')\rangle\langle \tilde{\phi}^* (t'')|+U^V(t',t''),\label{U3}
\end{eqnarray}
where  $U^V(t',t'')=\int dw|\psi^V_{w}(t')\rangle\langle\psi^V_{w}(t'')|$ is the Volkov-TEO. Using Eqs.~(\ref{exact}) and (\ref{U3}), we derive the SFA amplitude:
\begin{eqnarray}
m(p)&=&-i\int dt \langle \psi^V_{p}(t)|H_I(t)|\phi(t)\rangle\nonumber\\
 &&- i \int dt \langle \psi^V_p(t_f)|\tilde{\phi}^*(t_f)\rangle\langle\tilde{\phi}^*(t)|H_I(t)|\phi(t)\rangle.
\end{eqnarray}
Further, we neglect direct bound-bound transitions during the laser dressing of the bound state, assuming that the  laser-dressed  bound state emerges from the corresponding bare bound state due to the action of the Volkov-Dyson expansion:
\begin{eqnarray}
&&|\tilde{\phi}^*(t)\rangle\approx -i \int^t dt' U^V(t,t')VU^V(t',t_i) |\phi^*(t_i)\rangle,\nonumber\\
&&\langle\tilde{\phi}^*(t)|\approx -i \int_t dt'\langle\phi^*(t_f)| U^V(t_f,t')VU^V(t',t),
\label{dress}
\end{eqnarray}
where $t_i$ is  the initial time  when the laser field is turned on, and $|\phi^*(t)\rangle$ is the corresponding bare atomic eigenstate. The potential $V$ is accounted for perturbatively to describe  recombination to the excited state during the sub-barrier rescattering. Here it is assumed that the zeroth order term yields an unphysical boundary term and is neglected. Hence the amplitude reads
\begin{eqnarray}
m(p)&=& m_1(p)+m_3(p)\label{U4}\\
m_1(p)&=&-i \int dt \langle \psi^V_{p}(t)|H_I(t)|\phi(t)\rangle\label{mD}\\
 m_3(p)&=&i\int dt\int_{t} dt'\int dt'' \int dq\int dw\int dv\nonumber\\&&\times
 \langle \psi^V_{p}(t'')|V|\psi^V_v(t'')\rangle \langle\psi^V_{v}(t_i)|\phi^*(t_i)\rangle\langle\phi^*(t_f)|\psi^V_{w}(t_f)\rangle\nonumber\\
&&\times\langle\psi^V_{w}(t')|V|\psi^V_{q}(t')\rangle\langle\psi^V_{q}(t)|H_I(t)|\phi(t)\rangle,\label{mR}
\end{eqnarray}
with the direct ionization amplitude $m_1(p)$, and the ionization amplitude with a Freeman resonance $ m_3(p)$. To simplify the calculation of the high-order amplitude $ m_3(p)$, we model the atom by a short-range potential.

Dressing of the bound states emerges in Eq.~(\ref{U4}) due to transitions to intermediate Volkov states given by the integrations over the momenta $v$ and $w$. The mathematical structure of the dressing of the excited state  consists of two integrals of the form
\begin{eqnarray}
&&\int dx\int dp\exp\left[-i\int^{t}_{t_i}ds(p+A(s))^2/2-ipx-\kappa^* |x|\right]\nonumber\\&&=2\pi\exp\left[-i\int^t_{t_i}ds (i\kappa^*+(-1)^kA(s))^2/2\right],
\end{eqnarray}
which were  solved with the two-dimensional SPA. Here, the factor $(-1)^k$  corresponds to the $k^{th}$ half-cycle and arises from the derivative of $|\alpha(t)|$, with the excursion coordinate $ \alpha(t)=\int^t A(s) ds$. The integration over the momenta $v$ and $w$  results in dressing of the excited state with the vector potential of the laser field  via $i\kappa^* \rightarrow i\kappa^*+A(s)$,  effectively shifting the excited state energy  from $-I_p^*$ to $-I_p^*+U_p$, with the laser ponderomotive potential $U_p=E_0^2/4\omega^2$, $ \kappa^*=\sqrt{2I_p^*}$, and the ionization potential of the excited state $I_p^*$. The remaining integrals in $m(p)$ over the times $t$, $t'$, $t''$ and $q$ are calculated by SPA.

We gain a physical insight of the excitation process from the saddle-point conditions.
The saddle points for these four variables $t$, $t'$, $t''$ and $q$ are determined from the following equations
\begin{eqnarray}
[q(t',t)+A(t)]^2/2&=&-\kappa^2/2 \label{SPA1}\\
 q(t',t)&=&-\frac{\alpha(t')-\alpha(t)}{t'-t}\label{SPA2}\\
(i\kappa^{*}+(-1)^j A(t'))^2/2 &=& (q(t',t)+(-1)^j A(t'))^2/2 \label{SPA3}\\
(p+(-1)^i A(t''))^2/2&=& (i\kappa^{*}+(-1)^i A(t''))^2/2,\label{SPA4}
\end{eqnarray}
with the factors $(-1)^i$ and $(-1)^j$ corresponding to the processes in the $i^{th}$ and $j^{th}$ half-cycles. The Eqs.~(\ref{SPA1})-(\ref{SPA4}) describe the electron dynamics during the Freeman resonance. The ionization path begins at $t$ from the ground state, see Eq.~(\ref{SPA1}). The electron revisits the atomic core at   $t'$, when the intermediate momentum $q(t',t)$ fulfills Eq.~(\ref{SPA2}), which may lead to the electron recombination into the dressed excited state, see Eq.~(\ref{SPA3}). We choose the recolliding trajectory with sub-barrier excursion dynamics. In this case all saddle points are complex with similar real parts, but different imaginary parts, and the interpretation of the sub-barrier motion is in order.  In the considered nonadiabatic tunneling regime the electron gains energy during tunneling which allows for the transition to the dressed excited state, according to Eq.~(\ref{SPA3}).  Finally, the electron is ionized from the dressed excited state  at time $t''$, given by Eq.~(\ref{SPA4}).

\begin{figure}
   \begin{center}
  \includegraphics[width=0.45\textwidth]{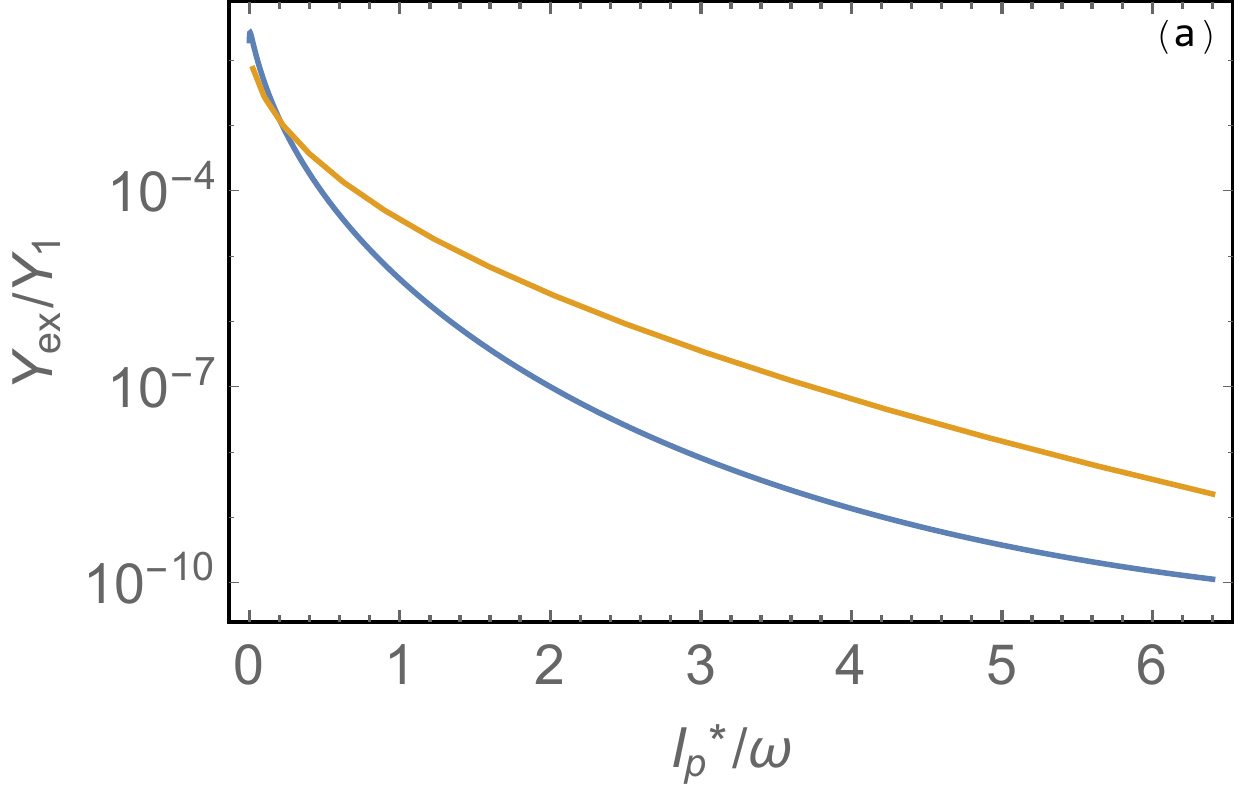}
\includegraphics[width=0.45\textwidth]{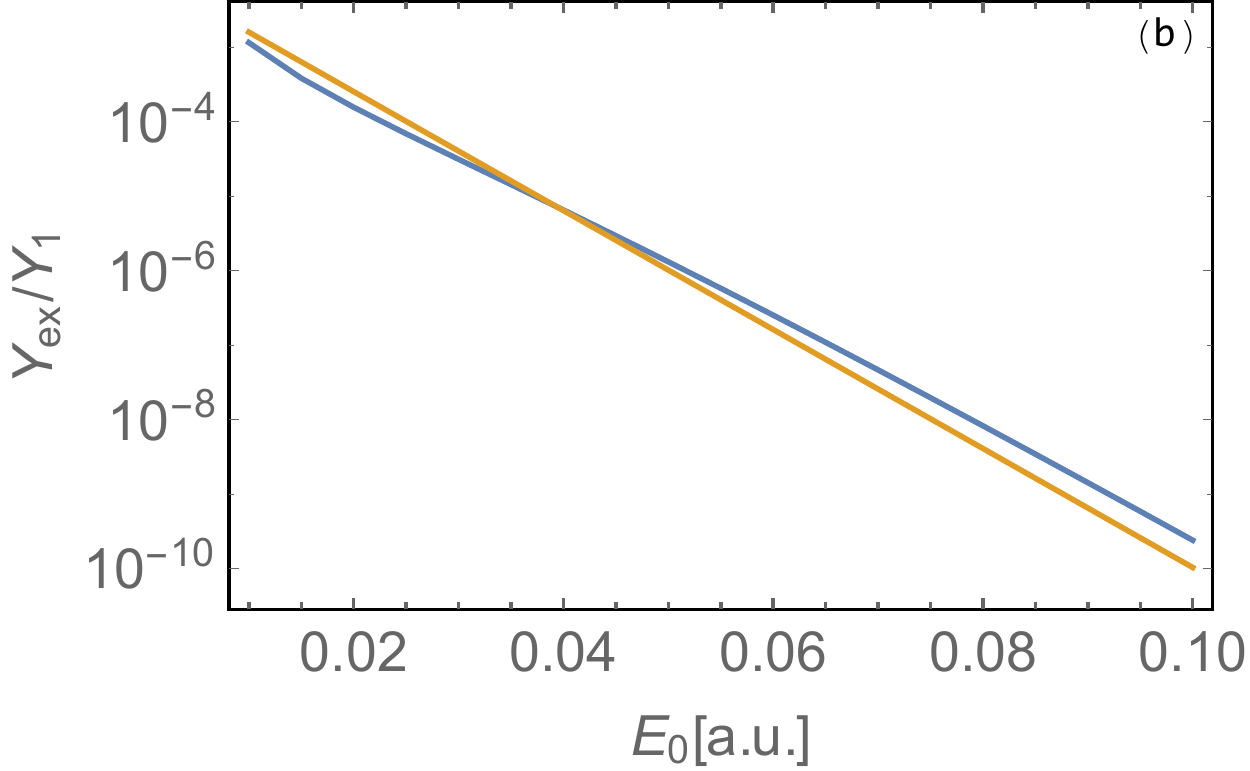}
\includegraphics[width=0.45\textwidth]{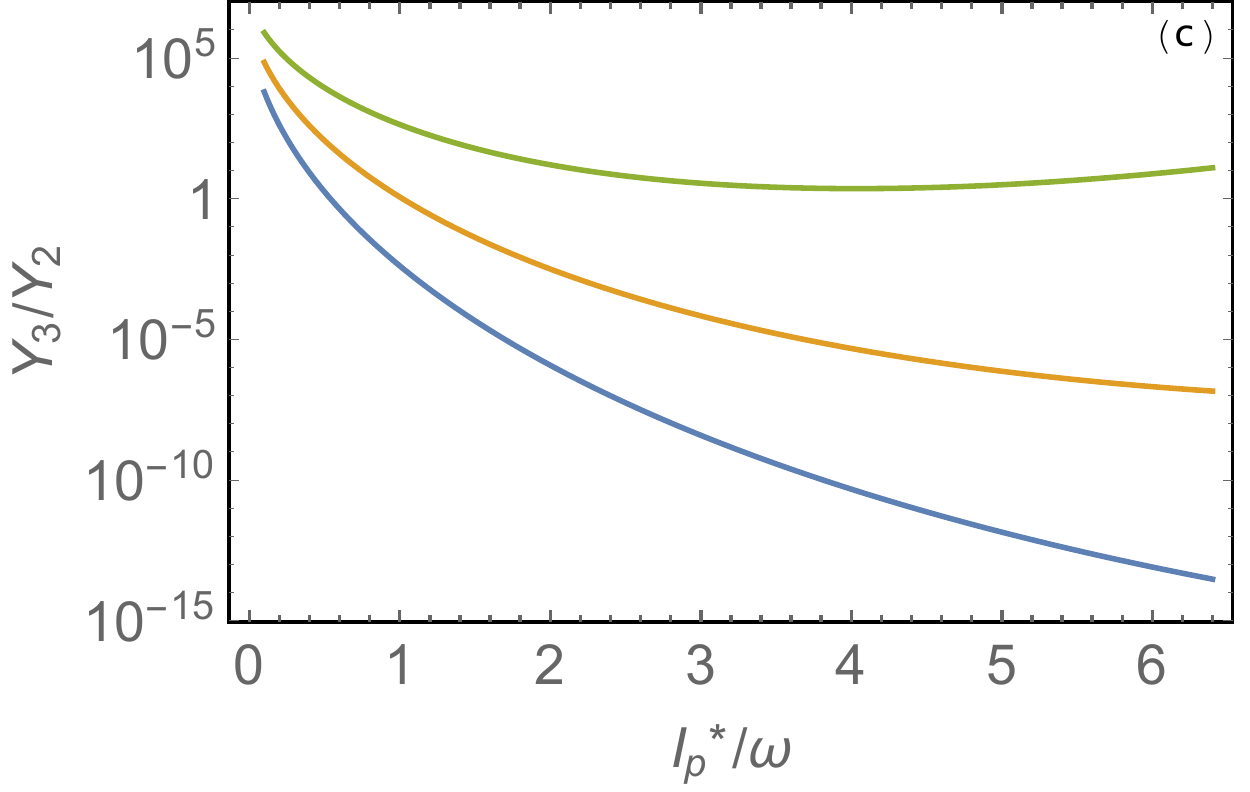}
 \caption{Ratio of the excitation yield  to that of the direct ionization $Y_{ex}/Y_1$ (blue) from a single half-cycle: (a) vs $I_p^*/\omega$, for $E_0=0.025$  a.u.; (b) vs the laser field amplitude $E_0 $ for $\kappa^*=0.23$ a.u.; $\omega=0.05$ a.u., $\kappa=1$ a.u., $\gamma=2$,  (orange) the scaling $\sim \exp(-2\kappa^* \alpha)$.  (c)
Ratio $Y_3/Y_2$ of the ionization yields due to the sub-barrier recollision with ($Y_3$) and without ($Y_2$) excited state, respectively,  for $E_0=0.035$ a.u. (blue), $E_0= 0.025$ a.u. (orange) and $E_0=0.015$ a.u. (green).}
         \label{m3-single-cycle}
    \end{center}
  \end{figure}

\section{Results}\subsection{The half-cycle contribution to the yield}\label{sec:half-cycle}

Firstly, we examine the excitation of a Rydberg state via sub-barrier recollision. Let us  analyze the contribution to the excitation yield $Y_{ex}$ during a half-cycle of the laser field. We define $Y_{ex}\equiv\int |m_{ex}(p)|^2dp$, with the excitation amplitude due to sub-barrier rescattering
 \begin{eqnarray}
m_{ex}&=&i\int dt\int_{t} dt'  \int dq\int dw  \langle\phi^*(t_f)|\psi^V_{w}(t_f)\rangle\nonumber\\
&&\times\langle\psi^V_{w}(t')|V|\psi^V_{q}(t')\rangle\langle\psi^V_{q}(t)|H_I(t)|\phi(t)\rangle,\label{m_ex}
\end{eqnarray}
which is derived from Eq.~(\ref{mR}), dropping the amplitudes of ionization from the excited state. The ratio of the resonant excitation  to the direct ionization yield, $Y_{ex}/Y_1$, is shown in Figs.~\ref{m3-single-cycle}(a) and (b), where $Y_1=\int  |m_1(p)|^2 dp$. We see that the excitation during a half-cycle is quite small.
The excitation probability is significantly damped at large $I_p^*$, and at large fields, which can be explained as follows. The process takes place at the laser field maximum, when the spatial distribution of the dressed excited state is concentrated at the distance $\alpha\sim E_0/\omega^2$ away from the core, with the width $\sim 1/\kappa^*$, meanwhile, for recombination the recolliding electron arrives at the core,   because momentum transfer from the core is needed for recombination. As a results  the recombination into the excited state is suppressed by a factor $\exp(-2\kappa^ {*}\alpha)$, see Figs.~\ref{m3-single-cycle}(a) and (b).

How the availability of the intermediate excited state changes the probability of the sub-barrier path is demonstrated in Fig.~\ref{m3-single-cycle}(c), where the ratio $Y_3/Y_2$ of the ionization yield during an half-cycle period due to the sub-barrier recollision with ($Y_3=\int |m_3|^2 dp$) and without ($Y_2=\int |m_2|^2dp$) excited state is shown. Here, the sub-barrier recollision is described in the second-order SFA by the matrix element
\begin{eqnarray}
 m_2=-\int dt\int_{t} dt'  \int dq\langle\psi^V_{p}(t')|V|\psi^V_{q}(t')\rangle\langle\psi^V_{q}(t)|H_I(t)|\phi(t)\rangle.\nonumber
\end{eqnarray}
During the half-cycle there is no resonance enhancement in the excitation, which is created only due to multi-cycle interference. Nevertheless, we see that even in that case the intermediate bound state increases the ionization probability several times at  small energies of the excited state $I_p^*$ and weak fields. There are two reasons for this enhancement. Firstly, the electron gains energy during the dwelling in the excited state, which increases further ionization probability. For instance, the energy of the electron at the recombination $t'$ is approximately $-0.3$ a.u. at $\omega=0.05$ a.u. for all values of the Keldysh parameter $\gamma=\kappa\omega/E_0$, whereas the laser-dressed energy at the tunneling from  the excited state at $t''$ is significantly larger, approximately $-I^*_p$ for small final momenta. Secondly, the electron wave packet spreading is suppressed during the dwelling time in the excited state. The spreading factor is  dominating,  and it is larger for larger Keldysh parameters, therefore the enhancement due to absence of spreading is also larger for larger $\gamma$. At large fields and large ionization energies of the excited state the recombination into the excited state is suppressed by the factor $\sim \exp(-2\kappa^* \alpha)$, which suppress severely the excitation probability and the ionization via Freeman resonances. We note that the  enhancement of the ionization yield due to  the transient excitation  is not large in the half-cycle contribution, but significantly boosted due to multi-cycle interference, as discussed below.

\subsection{Multi-cycle interference}\label{sec:multi-cycle}

\begin{figure}
   \begin{center}
\includegraphics[width=0.5\textwidth]{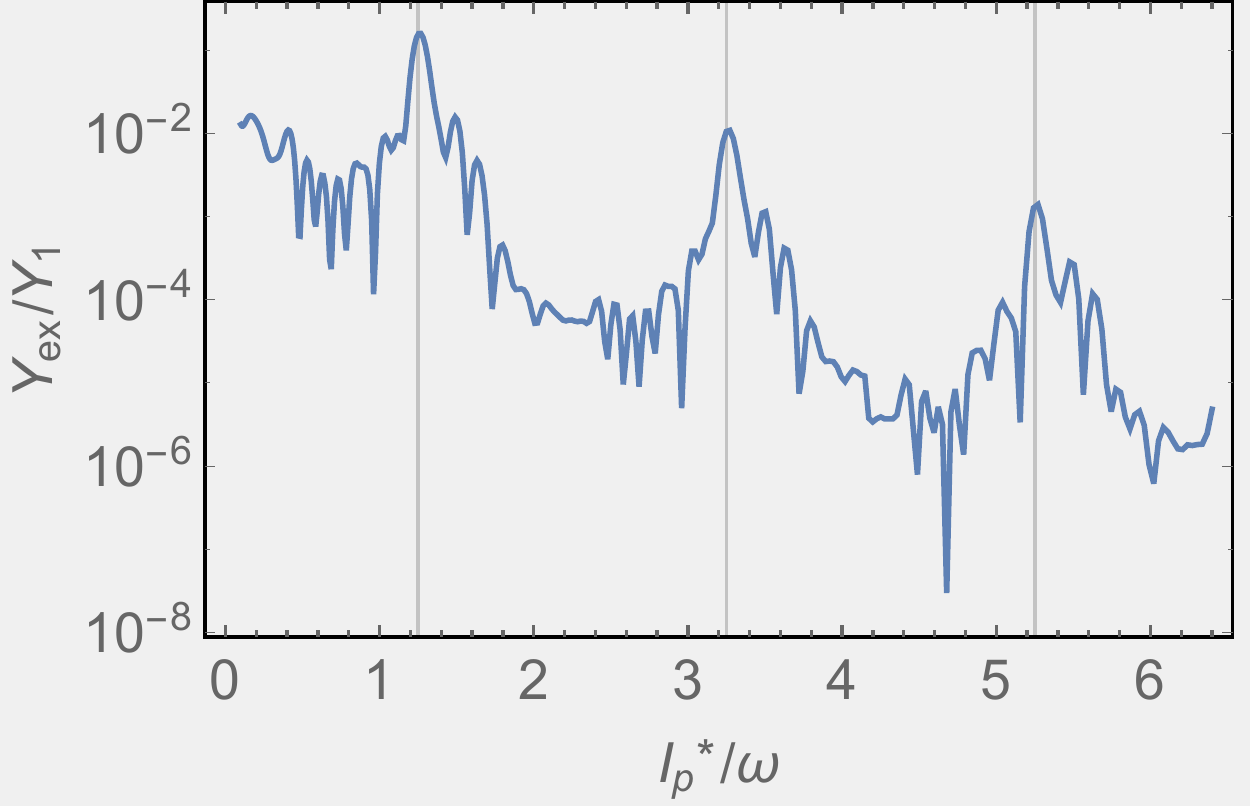}
 \caption{ Ratio of the excitation yield  to that of the direct ionization $Y_{ex}/Y_1$ from 10 cycles,  for $E_0=0.025$ a.u, $\omega=0.05$ a.u., ($\gamma=2$ for the ground state $\kappa=1$). The resonance conditions $ (U_p+I_p^*-I_p)=2\ell\omega$, with an integer $\ell$,  for the given excited state (${\cal P}_e=1$) are indicated by vertical lines.}
        \label{m3-5-half-cycle}
    \end{center}
  \end{figure}

A conspicuous resonance effect emerges in a long laser pulse when interference of contributions in the ionization amplitudes from different laser cycles is included. The structure of the $m_3$ amplitude is a product of an amplitude of direct ionization and a recombination amplitude into the dressed excited state. The three-dimensional saddle points are, therefore, decoupled and recombination can happen in a different half-cycle than the direct ionization. Consequently we add:
\begin{eqnarray}
m_3=\sum_i m^*_{1,i}\sum_{j\leq i} m_{ex,j}
\label{sum}
\end{eqnarray}
where the  ionization from the excited state takes place in the $i^{th}$ half-cycle, and the recombination in the $j^{th}$ one, where the time ordering ionization  after recombination is insured by $j\leq i$.
The phase of the excitation amplitude has the form
\begin{eqnarray}
\Phi_j&=&(-1)^j  \kappa^*\left[\alpha(t')-\alpha(t_i)\right]-\left[\frac{\kappa^{*2}}{2}t''-\beta(t'')+\beta(t_i)\right]\\
&&-\int^{t'}_{t''}ds\frac{\left[q(t',t'')+A(s)\right]^2}{2}+ \frac{\kappa^2}{2}t''+\frac{1-{\cal P}_e}{2}i\log[\alpha(t)],\nonumber
\label{phase}
\end{eqnarray}
where $j$ is the half-cycle number, $\dot{\beta}(t)=A^2(t)/2$, and ${\cal P}_e$ is the parity of the excited state. Two consecutive half-cycles with $t'\rightarrow t'+\pi/\omega$ and $t'' \rightarrow t''+\pi/\omega$ have  the phase difference $\Delta\Phi=\pi(U_p+I_p^*-I_p)/\omega+\pi (1-{\cal P}_e)/2$.  The interference of excitation amplitudes in the second sum of Eq.~(\ref{sum}) is constructive, and the yield is enhanced, if the resonance condition is fulfilled:
\begin{eqnarray}
U_p+I_p^*-I_p=\ell\omega ,
\label{resonance}
\end{eqnarray}
with an integer $\ell$; even $\ell$ corresponds to the case of the same parity of the ground and the excited states, and odd $\ell$ to the opposite parities. For the given excited state, the resonance peaks with respect to the state energy (or the laser intensity) have $2\omega$ separation, see Fig.~\ref{m3-5-half-cycle}. It is  due to constructive interference of the excitation amplitudes originating from each half-cycle. Comparison of the multi-cycle yield in Fig.~\ref{m3-5-half-cycle} with that of a single-half cycle one of Fig.~\ref{m3-single-cycle}(a) shows that the yield  $Y_{ex}/Y_1$ scales roughly quadratically with the number of laser cycles. This is because the resonance comes from the coherent contributions of half-cycle terms in the sum $\sum_{j} m_{ex,j}$ in Eq.~(\ref{sum}). Our model includes only one  (lowest) excited state, more resonances are possible when  higher excited states are considered.

\begin{figure}
   \begin{center}
   \includegraphics[width=0.5\textwidth]{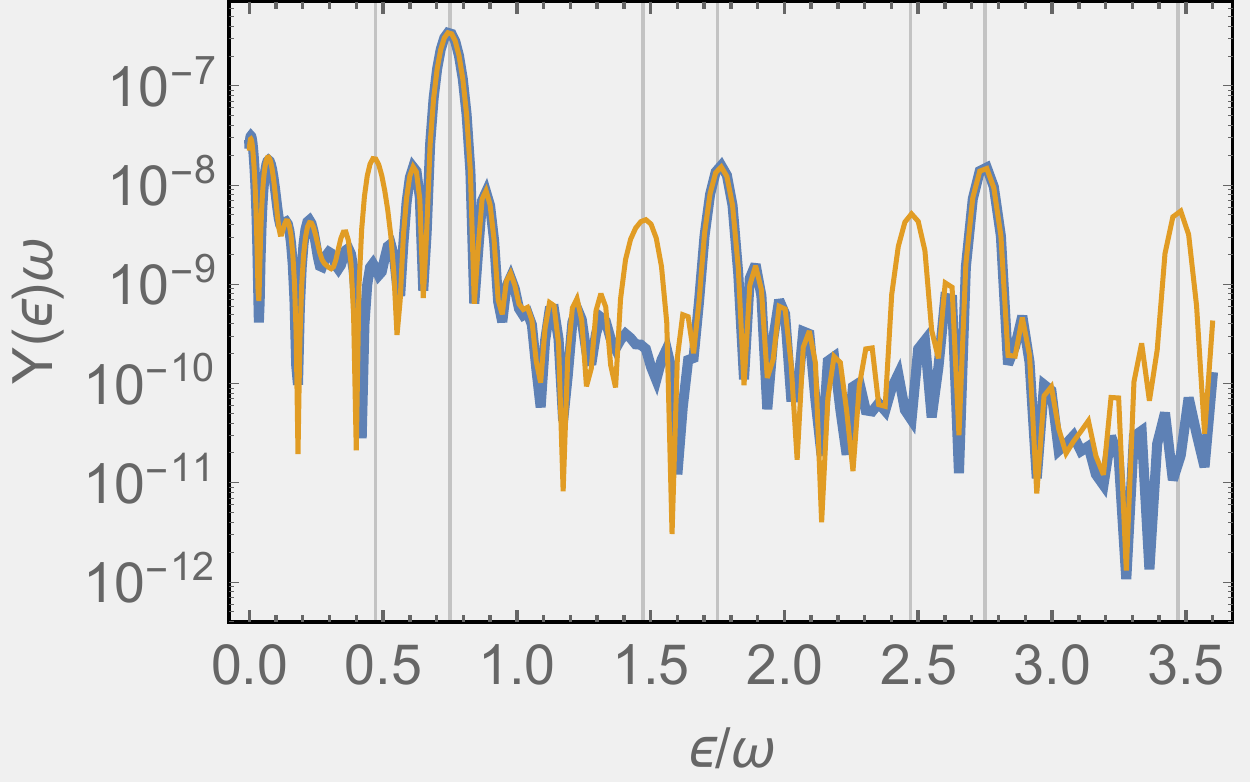}
 \caption{Photoelectron energy spectrum in a 10-cycle laser pulse in a logarithmic scale: (blue) direct ionization via $m_1$ and (orange) ionization at a Freeman resonance via $m_1+m_3$; $E_0=0.025$ a.u., $\omega=0.05$ a.u., $\kappa^*=0.23$ a.u., $\kappa=1$ a.u. $\gamma=2$. Grid lines indicate the ionization from the ground state $n-(I_p+U_p)/\omega$, and from the excited state $n-(I^*_p+U_p)/\omega$; $U_p+I^*_p-I_p=10.72$.}
       \label{m3-8-cycle}
    \end{center}
  \end{figure}

\subsection{Photoelectron spectra}\label{sec:spectra}

We analyze the signature of the resonant transient excitation during strong field ionization in the photoelectron spectra. Photoelectron energy distribution  in  a 10-cycle laser pulse is shown in Fig.~\ref{m3-8-cycle} in the case of the resonant excitation. We see that our model of excitations via sub-barrier recollision is able to describe the typical photoelectron spectra at Freeman resonances. Double peak structures arises in spectrum, see Fig.~\ref{m3-8-cycle}. One peak in series corresponds to the  direct ionization from the ground state with the energy conservation  $n\omega=p^2/2+U_p+I_p$. The second peak in series is due to the Freeman resonance. It corresponds to the multiphoton transition from the excited state with the energy $-I_p^*+U_p$ to the continuum with the energy $p^2/2+U_p$, with the energy conservation  $n\omega=p^2/2+I_p^*$, see grid lines in Fig.~\ref{m3-8-cycle}.

\section{Conclusion}\label{sec:conclusion}

We have proposed an intuitive model for Freeman resonances in the nonadiabatic tunneling ionization. Using specific quantum orbits provided by strong field approximation theory, we show a concrete pathway leading to the  transient population of  intermediate excited states during strong field ionization.  What is interesting, the pathway along which the electron gains energy necessary for the transfer to the excited state,  mostly travels under-the barrier, reflects from the outer surface of the barrier, propagates back to the core, and recombines to the excited bound state. All this sub-barrier  recollision takes place during imaginary time within a single half-cycle, visualizing the electron vertical transition in the strongly driven atom \cite{Ivanov_2005}.  We found that the available excited bound state can increase the probability of the sub-barrier recolliding pathway, even during the singe half-cycle contribution. The latter is mostly due to the suppressed spreading of the electron wave packet, during the dwelling time in the excited state. Although the transition probability during an half-cycle is small, it is resonantly enhanced due to constructive interference of contributions to the excitation amplitude emerging from different half-laser cycles, proportional to the square of the number of half cycles. As each half-cycle gives an interfering contribution, the resonance condition of different orders  for a ceratin excited state in this model are separated by twice of a photon energy.  The described sub-barrier pathway of the Freeman resonance is relevant  in the nonadiabatic tunneling regime, when the electron gains energy during tunneling, enabling transition to the laser dressed excited state.

 As an outlook beyond the scope of this paper, let us note on a possible application of the presented model of the Freeman resonances via sub-barrier recollision on the strong field electron-positron pair production problem in ultrastrong laser fields. When pairs are produced during the impinging of the laser beam on a nucleus (ion, or other atomic system), then the produced electron from vacuum due to multiphoton process can be captured into the bound state in the Coulomb potential. This bound-free channel of pair production has been thoroughly investigated in Refs.~\cite{Mueller_2003,Mueller_2004,Mueller_2008,Deneke_2008,Sommerfeldt_2019}. However, rather than real bound-free pair production, the state of a bound-free pair can emerge virtually as a transition state, which finally may end up with  a free electron and positron state. This pathway resembles conceptually to the Freeman resonance discussed in this paper. The present model of the under-the barrier recollision can be extended for the solution of Dirac equation describing excitation of the Dirac see electron into a positive energy state, with transient capture into the bound state.

\bibliography{strong_fields_bibliography}

\end{document}